\documentclass[prb,twocolumn,groupedaddress,showpacs]{revtex4}
\usepackage{graphicx}
\begin{document}
\title{Lead clusters: different potentials, different structures}
\author{Jonathan P.~K.~Doye}
\email{jpkd1@cam.ac.uk}
\affiliation{University Chemical Laboratory, Lensfield Road, Cambridge CB2 1EW, United Kingdom}
\date{\today}
\begin{abstract}
The lowest-energy structures of lead clusters interacting via a Gupta potential
are obtained for $N\le 150$. Structures based on Marks decahedra dominate at
the larger sizes. 
These results are very different from those obtained previously using 
a lead glue potential, and the origins of the differences 
are related back to differences
in the potential.
\end{abstract}
\pacs{61.46.+w,36.40.Mr}
\maketitle

\section{Introduction}
\label{sect:intro}
The structure of a metal cluster is one of its most important properties,
yet it is a property that can be hard to access for free 
clusters. Experiments usually lead only to an indirect measurement of
the structure, and so results need to be compared to what would be expected
for candidate structures. Therefore, there is an important need for 
reliable predictions from theory. However, this is also something that can 
be hard to achieve.

For all but quite small clusters, performing a full search for the 
lowest-energy structure is unfeasible with {\it ab initio} electronic 
structure methods. Therefore, one needs to use some kind of potential.
However, the energy differences between competing structures can often
be relatively small. Furthermore, for a cluster's energy to be accurate, 
the energetics associated with a wide range of properties potentially 
need to be well modelled---different external surfaces, twin planes,
different crystal structures, and the response to strain.
To model the potential temperature dependence of the structure 
vibrational properties need also to be well described.\cite{Doye01b}
Therefore, prediction of the correct structure of a cluster represents
a tough challenge for a potential.
Furthermore, often it is not clear which of the available potentials for a 
system will work best.

It is my contention that an important aspect of assessing a potential's 
quality for modelling clusters is to first understand the relationship
between the observed structures and the potential.
As well as providing fundamental insights into cluster structure, 
this can help to determine whether the observed structures reflect some 
robust feature of the potential, or rather whether they stem from some 
deficiencies in the potential. This is the approach I undertake here.
I first find the lowest-energy structures for lead clusters modelled using
a Gupta potential, and then compare them to previous results for a lead
glue potential.\cite{Doye03a}
The origins of the differences are explored, 
highlighting the implications for the relative merits of the two potentials.

\begin{figure*}
\begin{center}
\includegraphics[width=14.5cm]{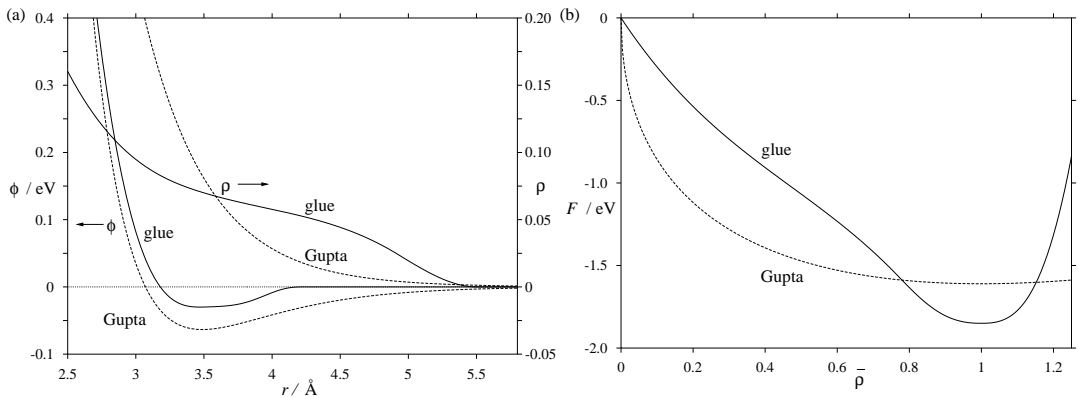}
\end{center}
\caption{(a) $\phi(r)$ and $\rho(r)$, and (b) $F(\bar\rho)$
for lead Gupta and glue potentials in the effective pair format.}
\label{fig:potential}
\end{figure*}

\section{Methods}
\label{sect:methods}

Both potentials that I consider are of the embedded-atom form,
where the potential energy is given by
\begin{eqnarray}
E&=&E_{\rm pair} + E_{\rm embed} \nonumber \\
 &=&\sum_{i<j} \phi\left(r_{ij}\right)+\sum_{i} F\left(\bar\rho_i\right)\label{eq:EAM}, 
\end{eqnarray}
where $\phi(r)$ is a short-ranged pair potential,
$F(\bar\rho)$ is a many-body embedding (or glue) function
and 
$\bar\rho_i=\sum_j \rho\left(r_{ij}\right)$,
where $\rho(r)$ is an ``atomic density'' function.
One potential is of the Gupta form:\cite{Cleri93}
\begin{eqnarray}
\phi\left(r\right)&=&2A e^{-p(r/r_0-1)}\nonumber\\
\label{eq:Gupta}
F\left(\bar\rho\right)&=&-\xi \sqrt{\bar\rho} \\
\rho\left(r\right)&=&e^{-2q(r/r_0-1)}.\nonumber
\end{eqnarray}
whereas for the glue potential there are no assumed forms for
these functions; instead they arise in the fitting process.\cite{Lim}
The parameters of the Gupta potential were fitted on the basis of bulk
properties,\cite{Cleri93} whereas a much wider variety of configurations, 
including surface properties, were used to parameterize the glue 
potential.\cite{Lim} 

However, the functions in Eq.\ \ref{eq:Gupta} are non-unique.
Functions that give rise to exactly the same energy can be
constructed by the transformation
\begin{eqnarray}
\phi'\left(r\right)&=&\phi\left(r\right) + 2 g \rho(r) \nonumber\\
F'\left(\bar\rho\right)&=&F\left(\bar\rho\right) - g \bar\rho ,
\end{eqnarray}
This transformation redistributes the total energy between
$E_{\rm pair}$ and $E_{\rm embed}$.
When
\begin{equation}
g=\left.{dF\over d\bar\rho}\right|_{\bar\rho=\bar\rho_{\rm xtal}},
\end{equation}
$F'(\bar\rho)$ has a minimum at $\bar\rho_{\rm xtal}$,
where $\bar\rho_{\rm xtal}$ is the value of $\bar\rho$ in the equilibrium crystal.
This choice is called the effective pair format, and
has been suggested as the most natural way to partition the
energy between the pair and many-body contributions.\cite{Johnson89}
In this format, when $\bar\rho=\bar\rho_{\rm xtal}$ the
pair potential controls the energy change for any change of configuration
that does not significantly alter $\bar\rho$. 
This form also makes comparisons between different potentials easier.

Indeed, the components of the two lead potentials are 
compared in Fig.\ \ref{fig:potential},
and are quite different. The pair potential contributes less
for the glue potential. At its minimum the energy is only 1.62\% of the minimum
in the embedding function, compared to 3.94\% for the Gupta form. The $\rho(r)$
functions are also very different. For the Gupta potential it decreases
rapidly and uniformly because of its exponential form, whereas the glue 
$\rho(r)$ has a wide plateau around the nearest-neighbour distance.

The optimal clusters will have the best balance between maximizing 
the pair and embedding energies. As is well-understood,\cite{Doye95c}
maximization of the pair energy is achieved through a balance
between maximizing the number of bonds (i.e.\ spherical and facets with
high coordinate surface atoms, e.g.\ fcc $\{111\}$), and minimizing the
strain energy (the energetic penalty for distances that deviate from 
the minimum of the potential).

\begin{figure*}
\begin{center}
\includegraphics[width=13cm]{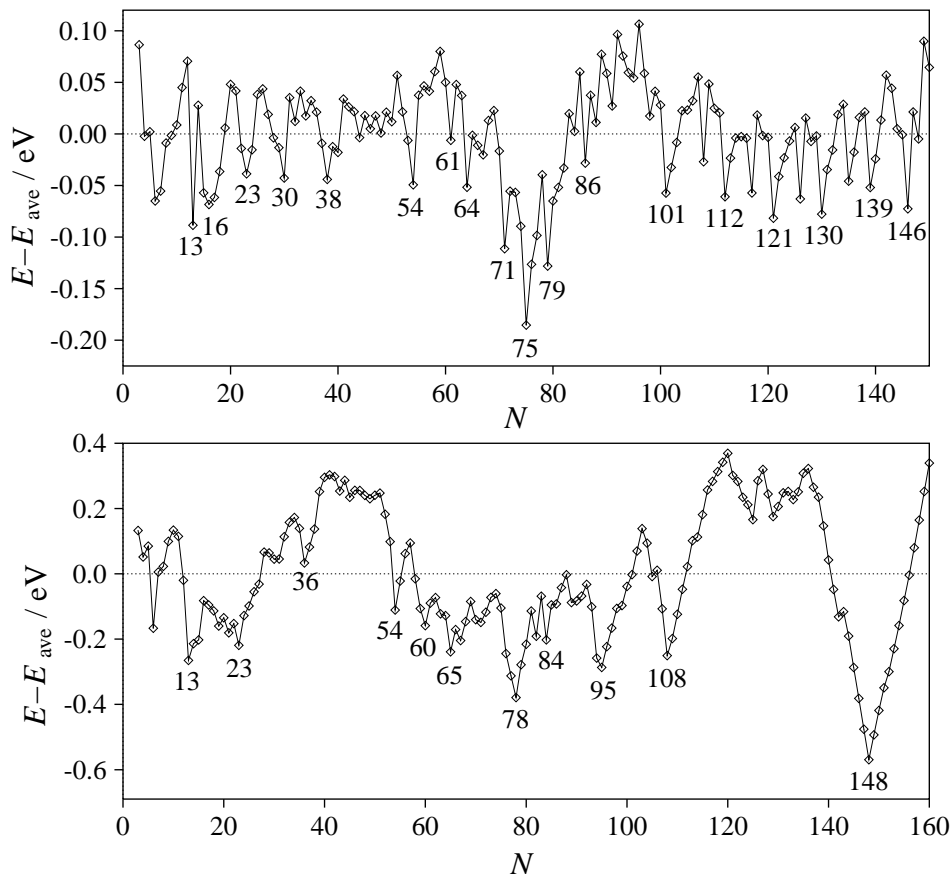}
\end{center}
\caption{Energies of the lead clusters modelled by the (a) Gupta and (b) 
glue potentials relative to $E_{\rm ave}$, a four-parameter fit to the
energies. $E_{\rm ave}^{\rm Gupta}=-2.0631 N+0.9778 N^{2/3}-
1.0084 N^{1/3}+1.2917$;  $E_{\rm ave}^{\rm glue}=-2.0251 N+1.6608 N^{2/3}+
1.3662 N^{1/3}-0.8634$.}
\label{fig:EvN}
\end{figure*}

The way the embedding energy is maximized and the effects on structure
have only begun to be explored more 
recently.\cite{Doye03a,Michaelian99,Soler00,Soler01,Baletto02b,Doye03d,Doye03e}
For $E_{\rm embed}$ to be maximal the $\bar\rho_i$ for each atom need to 
be as close to $\bar\rho_{\rm xtal}$ as possible. An important feature is that
an atom with a low (high) coordination number can compensate for this by 
shortening (extending) the bonds. Therefore, the embedding energy will 
favour a shrinking of the surface, which can in turn compress the 
interior of the cluster. This can be problematic for structures such 
as the Mackay icosahedron, where bond distances for surface atoms are 
longer than interior atoms, and so surface compression can lead to
unfavourable energies for the atoms at the core. 
The magnitude of these effects will depend on the relative depths
of the effective pair and embedding functions, as the pair potential
will tend to resist any changes in structure that lead to
distances deviating from the preferred pair distance.
When the embedding energy dominates unusual structures that naturally
have bond distances that are shorter on the surface can be 
favoured.\cite{Doye03d,Doye03e}

For Gupta potentials the value of $p/q$ provides an approximate indicator 
of the relative weights of the pair and embedding energies. As
$p$ tends to $2q$ the minimum in the effective pair potential 
disappears.\cite{Doye03e} Therefore, one tends to see a progression from
potentials that favour icosahedra to decahedra to close-packed clusters
and finally disordered clusters as $p\rightarrow 2q$.
This is somewhat simplistic since as one goes further away from $p=2q$ the
other parameters of the potential play an increasing role in the relative
depths of the pair and embedding functions.

Global optimization of the lead clusters to locate the lowest-energy
structures was achieved using the basin-hopping method,\cite{WalesD97}
supplemented by reoptimizations of databases of structures found previously 
for other clusters.\cite{web}

\begin{table*}
\caption{Energies and point groups of the global minima for lead clusters
modelled using a Gupta potential. For $N\ge 13$ a structural assignment
has been made, where i stands for icosahedral, d for decahedral, 
fcc for face-centred cubic, cp for close-packed, dd for distorted decahedral
and pt for polytetrahedral. If no assignment has been given the cluster
has no apparent overall order.
\label{tab:E}}
\begin{ruledtabular}
\begin{tabular}{ccccccccccccccccccc}
$N$ & PG & Energy & & & $N$ & PG & Energy & & & $N$ & PG & Energy & & & $N$ & PG & Energy & \\
\hline
   3 &  $D_{3h}$ &    -4.231623 & & &   40 &  $D_2$  &   -73.262766 & & &   77 &  $C_{2v}$ &  -144.258747 & d & &  114 &  $C_s$    &  -215.807472 & d \\ 
   4 &  $T_d$    &    -6.099714 & & &   41 &  $C_1$    &   -75.112816 & & &   78 &  $C_s$    &  -146.128629 & d & &  115 &  $C_s$    &  -217.749269 & d \\ 
   5 &  $D_{3h}$ &    -7.886887 & & &   42 &  $C_s$    &   -77.023173 & & &   79 &  $O_h$    &  -148.146432 & fcc & &  116 &  $C_{3v}$ &  -219.694133 & fcc \\ 
   6 &  $O_h$    &    -9.755691 & & &   43 &  $C_{2v}$ &   -78.931871 & d & &   80 &  $C_{4v}$    &  -150.013008 & fcc & &  117 &  $C_s$    &  -221.691036 & d \\ 
   7 &  $D_{5h}$ &   -11.556399 & & &   44 &  $C_2$    &   -80.861958 & & &   81 &  $C_{2v}$ &  -151.930037 & d & &  118 &  $C_s$    &  -223.559459 & d \\ 
   8 &  $D_{2d}$ &   -13.327553 & & &   45 &  $C_1$    &   -82.746570 & & &   82 &  $C_{2v}$ &  -153.841784 & d & &  119 &  $C_s$    &  -225.523689 & d \\ 
   9 &  $C_{2v}$ &   -15.144569 & & &   46 &  $C_3$    &   -84.666164 & & &   83 &  $C_{2v}$ &  -155.720227 & d & &  120 &  $C_s$    &  -227.470026 & d \\ 
  10 &  $C_{3v}$ &   -16.964625 & & &   47 &  $C_1$    &   -86.561578 & & &   84 &  $C_s$    &  -157.668905 & d & &  121 &  $C_{2v}$ &  -229.493351 & d \\ 
  11 &  $C_{2v}$ &   -18.763846 & & &   48 &  $C_1$    &   -88.486884 & & &   85 &  $C_s$    &  -159.543391 & d & &  122 &  $C_{2v}$ &  -231.398205 & d \\ 
  12 &  $C_{5v}$ &   -20.578513 & & &   49 &  $C_1$    &   -90.376393 & & &   86 &  $C_{3v}$ &  -161.564185 & fcc & &  123 &  $C_{2v}$ &  -233.325446 & d \\ 
  13 &  $I_h$    &   -22.582269 & i & &   50 &  $D_{3h}$ &   -92.296124 & cp & &   87 &  $C_s$    &  -163.431365 & fcc & &  124 &  $C_{2v}$ &  -235.254926 & d \\ 
  14 &  $C_{6v}$ &   -24.314590 & pt & &   51 &  $C_s$    &   -94.162526 & cp & &   88 &  $C_s$    &  -165.391034 & fcc & &  125 &  $C_{3v}$ &  -237.188016 & cp \\ 
  15 &  $D_{6d}$ &   -26.251716 & pt & &   52 &  $C_{2v}$ &   -96.110032 & & &   89 &  $C_s$    &  -167.258681 & d & &  126 &  $C_s$    &  -239.203573 & d \\ 
  16 &  $C_{2v}$ &   -28.118745 & dd & &   53 &  $C_{3v}$ &   -98.050794 & & &   90 &  $C_s$    &  -169.211406 & fcc & &  127 &  $C_s$    &  -241.071732 & d \\ 
  17 &  $C_{2v}$ &   -29.970808 & dd & &   54 &  $I_h$    &  -100.007540 & i & &   91 &  $C_{2v}$ &  -171.177663 & cp & &  128 &  $C_s$    &  -243.041253 & d \\ 
  18 &  $C_s$    &   -31.807535 & dd & &   55 &  $I_h$    &  -101.835280 & i & &   92 &  $C_s$    &  -173.043249 & cp & &  129 &  $C_s$    &  -244.983046 & d \\ 
  19 &  $C_s$    &   -33.630069 & dd & &   56 &  $D_{2h}$    &  -103.745751 & fcc & &   93 &  $C_1$    &  -174.999552 & cp & &  130 &  $C_{2v}$ &  -247.006258 & d \\ 
  20 &  $C_s$    &   -35.455114 & dd & &   57 &  $C_{2v}$ &  -105.662759 & d & &   94 &  $C_s$    &  -176.951443 & cp & &  131 &  $C_{2v}$ &  -248.910730 & d \\ 
  21 &  $C_{6v}$ &   -37.331154 & pt & &   58 &  $C_1$    &  -107.560437 & & &   95 &  $C_{2v}$ &  -178.892736 & d & &  132 &  $C_s$    &  -250.839600 & d \\ 
  22 &  $C_s$    &   -39.259153 & & &   59 &  $C_{2v}$ &  -109.458501 & d & &   96 &  $C_s$    &  -180.777213 & cp & &  133 &  $C_s$    &  -252.753620 & d \\ 
  23 &  $C_2$    &   -41.158169 & & &   60 &  $C_{2v}$ &  -111.406643 & d & &   97 &  $C_s$    &  -182.761992 & d & &  134 &  $C_1$    &  -254.691944 & d \\ 
  24 &  $C_s$    &   -43.011299 & & &   61 &  $C_{3v}$ &  -113.381653 & cp & &   98 &  $C_s$    &  -184.740639 & fcc & &  135 &  $C_s$    &  -256.715113 & d \\ 
  25 &  $C_1$    &   -44.836233 & & &   62 &  $C_s$    &  -115.247146 & cp & &   99 &  $C_s$    &  -186.654481 & d & &  136 &  $C_{4v}$ &  -258.635983 & fcc \\ 
  26 &  $C_s$    &   -46.711166 & & &   63 &  $C_s$    &  -117.178015 & cp & &  100 &  $C_{5v}$ &  -188.605832 & d & &  137 &  $C_s$    &  -260.551431 & d \\ 
  27 &  $C_s$    &   -48.618276 & & &   64 &  $C_{2v}$ &  -119.187860 & d & &  101 &  $D_{5h}$ &  -190.629737 & d & &  138 &  $C_s$    &  -262.495627 & d \\ 
  28 &  $C_1$    &   -50.525086 & & &   65 &  $C_{2v}$ &  -121.058816 & d & &  102 &  $D_{2h}$ &  -192.543687 & fcc & &  139 &  $C_{2v}$ &  -264.518644 & d \\ 
  29 &  $C_2$    &   -52.420240 & & &   66 &  $C_s$    &  -122.990751 & d & &  103 &  $C_{2v}$ &  -194.458602 & d & &  140 &  $O_h$    &  -266.440716 & fcc \\ 
  30 &  $C_{3v}$ &   -54.337045 & & &   67 &  $C_{2v}$ &  -124.922594 & d & &  104 &  $C_{2v}$ &  -196.367268 & fcc & &  141 &  $C_{2v}$ &  -268.353333 & d \\ 
  31 &  $C_3$    &   -56.148062 & & &   68 &  $C_{3v}$ &  -126.812603 & cp & &  105 &  $C_{2v}$ &  -198.306531 & d & &  142 &  $C_s$    &  -270.260176 & d \\ 
  32 &  $D_{2d}$ &   -58.061383 & & &   69 &  $C_1$    &  -128.726858 & d & &  106 &  $C_s$    &  -200.237783 & d & &  143 &  $C_{3v}$ &  -272.223375 & cp \\ 
  33 &  $C_1$    &   -59.923961 & & &   70 &  $C_s$    &  -130.690514 & d & &  107 &  $C_s$    &  -202.155252 & cp & &  144 &  $C_s$    &  -274.213617 & d \\ 
  34 &  $C_2$    &   -61.841131 & & &   71 &  $C_{2v}$ &  -132.710251 & d & &  108 &  $C_s$    &  -204.178302 & d & &  145 &  $C_{5v}$ &  -276.170441 & d \\ 
  35 &  $C_{2v}$ &   -63.721030 & & &   72 &  $C_s$    &  -134.580108 & d & &  109 &  $C_s$    &  -206.044130 & d & &  146 &  $D_{5h}$ &  -278.193661 & d \\ 
  36 &  $C_2$    &   -65.627949 & & &   73 &  $C_s$    &  -136.507283 & d & &  110 &  $C_s$    &  -208.009245 & d & &  147 &  $C_s$    &  -280.051292 & d \\ 
  37 &  $C_{2v}$ &   -67.555415 & dd & &   74 &  $C_{5v}$ &  -138.466765 & d & &  111 &  $C_s$    &  -209.955620 & d & &  148 &  $C_{2v}$ &  -282.029242 & d \\ 
  38 &  $O_h$    &   -69.488671 & f & &   75 &  $D_{5h}$ &  -140.489863 & d & &  112 &  $C_{2v}$ &  -211.979029 & d & &  149 &  $C_s$    &  -283.886484 & d \\ 
  39 &  $C_{4v}$    &   -71.356501 & f & &   76 &  $C_{2v}$ &  -142.358601 & d & &  113 &  $C_{2v}$ &  -213.884008 & d & &  150 &  $C_{2v}$ &  -285.864042 & d \\ 
\end{tabular}
\end{ruledtabular}
\end{table*}

\begin{figure*}
\begin{center}
\includegraphics[width=14.5cm]{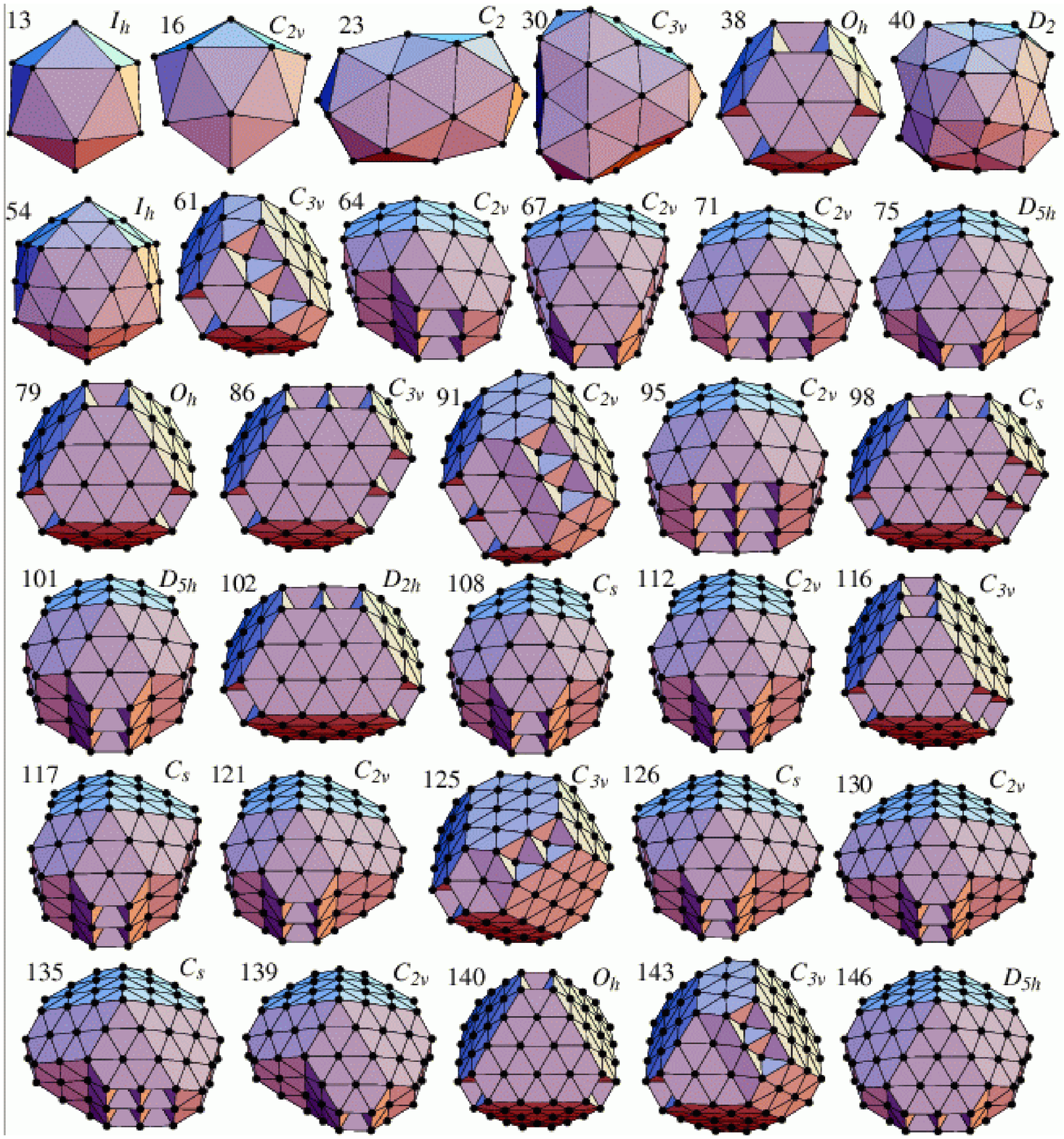}
\end{center}
\caption{A selection of clusters for lead clusters 
using the Gupta potential. Each structure is labelled by the size
and point group.}
\label{fig:Guptapics}
\end{figure*}

\section{Results}
\label{sect:results}

Putative global minima up to $N=150$ were found. 
For reference, their energies and point groups are given in Table \ref{tab:E}.
The coordinate files for all the global minima can be obtained online at
the Cambridge Cluster Database.\cite{web}
Results up to $N$=55 have previously been reported by Lai 
{\it et al.},\cite{Lai02} but the current results improve these
at $N$=43, 50, 51 and 54. The size-dependence of the
energies are plotted in Fig.\ \ref{fig:EvN}, and a selection
of clusters are depicted in Fig.\ \ref{fig:Guptapics}. 

Of the clusters for $N\ge 13$, 70 are decahedral, 29 are close-packed (both
perfect face-centred-cubic structures and those involving a twin plane), 6
have a distorted decahedral form recently characterized for zinc and
cadmium clusters\cite{Doye03e} (e.g.\ $N$=16), 
three are icosahedral and three have polytetrahedral structures.
The remaining 27 have no assigned structure, and are somewhat disordered.
As is fairly common, these `disordered' structures appear at the smaller sizes
in between the magic numbers at $N$=13, 38 and 54 that are associated 
with ordered clusters. Examples at $N$=23, 30 and 40 are shown in 
Fig.\ \ref{fig:Guptapics}.

Interestingly the uncentred Mackay icosahedron at $N$=54 is more stable than
the usual 55-atom Mackay icosahedron. The strong compressions at the centre
of the icosahedron resulting from shrinkage of the surface bonds makes
it unfavourable for this site to be occupied. This type of vacancy was
first noted by Boyer and Broughton for very large Lennard-Jones 
Mackay icosahedra,\cite{Boyer90} but is not uncommon for much smaller 
metal clusters because of many-body effects.

For $N\ge 59$ all the clusters are either decahedral or close-packed.
A significant proportion of the close-packed clusters
are in the region $N$=85--100, where decahedral clusters with five atoms 
along the fivefold axis 
are becoming less stable, yet those with six atoms
along this axis 
are not yet very competitive. It is noticeable that the 75-atom complete 
Marks decahedron is particularly stable (Fig.\ \ref{fig:EvN}(a)). However, 
the other complete Marks decahedra with 101 and 146 atoms are not as 
prominent in Fig.\ \ref{fig:EvN}(a) because of their rather prolate and
oblate structures, respectively. Indeed, a number of the intermediate 
asymmetric Marks decahedra have similar stabilities.

The observed structures tally quite well with the value of $p/q$ (2.625). 
There are fewer disordered and close-packed structures
than gold (e.g.\ $p/q=2.457$ (Ref.\ \onlinecite{Michaelian99})), 
but less icosahedral and more 
decahedral structures than for silver (e.g.\ 
$p/q=$3.003 (Ref.\ \onlinecite{Michaelian99})).\cite{Doye98c}
That is, the relative weights of the embedding and pair terms is such that
icosahedra are for the most part not favoured because of unfavourable 
embedding energies associated with compression of the cluster centre,
but decahedra are more prevalent than close-packed clusters because of 
a more favourable pair energy due to their greater proportion of $\{111\}$ 
faces.

\begin{figure*}
\begin{center}
\includegraphics[width=14.5cm]{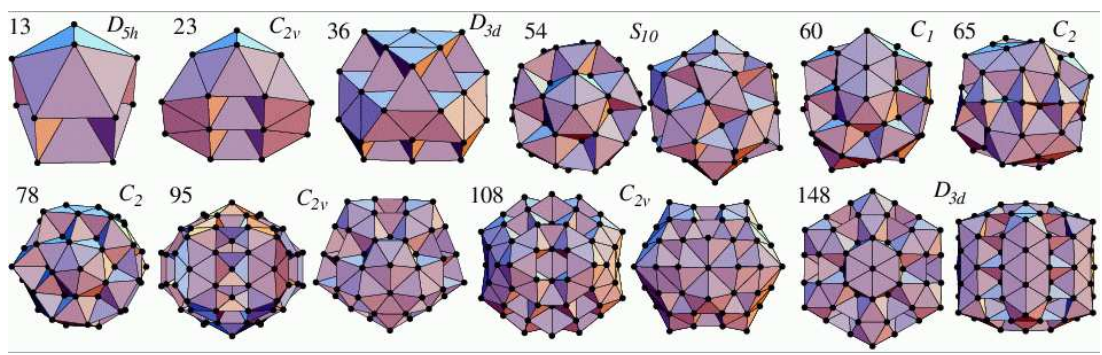}
\end{center}
\caption{A selection of particularly stable clusters for lead clusters 
using a glue potential.}
\label{fig:gluepics}
\end{figure*}

When these results are compared to those previously obtained for the 
glue potential\cite{Doye03a} there is virtually no overlap. The pattern of
magic numbers is completely different (Fig.\ \ref{fig:EvN}) and none of 
the structures are the same for $N\ge 8$. A selection of stable clusters
for the glue potential are depicted in Fig. \ref{fig:gluepics}
The clusters do not exhibit any of the usual structural forms, and
although most are fairly disordered, some strong magic numbers are apparent
in Fig.\ \ref{fig:EvN}(b). The strongest of these is from the highly 
symmetrical 148-atom hexagonal barrel structure. 
Also, of interest is the 54-atom cluster
as it is related to the uncentred icosahedron favoured for the Gupta potential
by a series of twists to the outermost shell around one of the fivefold axes.

The long-range character of $\rho(r)$ is in part because the potential
reproduces the small surface energy difference between the 
fcc $\{111\}$ and $\{100\}$ faces of lead.\cite{Lim} Although an atom on a 
$\{100\}$ faces has one less nearest neighbour, this
is in part compensated because it has more next-neighbours at 
$\sqrt 2 r_0$ that make a significant contribution to $\rho(r)$. 
For the clusters this does not lead to $\{100\}$ facets but most
of the surfaces of the clusters involve some combination of squares and
triangles. 

That more disordered structures are favoured for the glue potential compared to
the Gupta potential is simply because of the greater dominance of the embedding
term. However, the reasons why the glue global minima have the 
best embedding energy is more subtle.
In particular, it is related to the position of the cutoff, which
occurs between the second and third neighbour shells of the ordered structures.
By comparison the disordered clusters have a much broader spectrum of pair 
distances, and it is in fact the contributions to the $\bar\rho_i$ values
from distances near to the cutoff that tip the balance in favour of 
the observed structures.\cite{Doye03a} The position of this cutoff is not
physically motivated, but has been chosen for computational convenience.
As the structures do not seem to reflect robust features of the potential, 
they showed be viewed with appropriate skepticism.

Unfortunately, experiments on the structure of lead clusters are few. There
have been some mass spectroscopic studies of very small clusters,
but these are in the size range ($N<25$),\cite{Muhlbach82,LaiHing87} 
where empirical potentials are expected to be least applicable. The most
relevant experiments are 
those by Hyslop {\it et al.} using electron diffraction.\cite{Hyslop01} 
However, for the size range most relevant to the present study, 
no adequate fit to the diffraction patterns was possible, whereas 
the larger clusters appear to be dominated by decahedra. 

\section{Conclusions}
\label{sect:conc}

It is the contention of this paper that to make progress in predicting the
structure of clusters, understanding the relationship between the interactions
and the resulting structure is important. Such an approach will provide a 
better general framework for understanding the criteria for a potential to be 
successful, and specific insights into the utility of a particular 
potential.

That the two lead potentials we consider suggest a totally different pattern 
of structures illustrates some of the problems for 
predicting cluster structure. 
However, given the differences in the potentials when compared in their
effective pair format, the resulting structural 
differences are not in themselves so surprising.

The results also illustrate some of the pros and cons of using potentials
with a specified form as with the Gupta potential, versus those with
flexible forms. The former have less opportunity to produce manifestly 
unreasonable structures, but this is because of their more limited range
of structural behaviour. If the real structures lie outside of this
range, then such potentials have no chance of successful structural 
prediction. For example, the Gupta potential predicts that decahedral
structures dominate at the larger sizes in this study, whereas the
experimental electron diffraction suggest none of the standard forms.\cite{Hyslop01}

In contrast, flexible potentials have more chance of predicting some 
of the unusual 
structures that may occur at small sizes. However, there can be 
unforeseen consequences in regions of configuration space that lie 
outside of the possibilities considered in the fitting process.
In the current case, the lead glue clusters are able to exploit 
questionable features of the potential, and so the structures are 
probably not realistic. 
This is not because this potential is a bad one---it
has been widely used with reasonable success. Rather, it 
illustrates how predicting cluster structures is a very stringent test
of a potential.

\end{document}